\documentclass[12pt]{article}
\usepackage{amsmath}
\usepackage{amsfonts}
\usepackage{amssymb}

\providecommand{\U}[1]{\protect\rule{.1in}{.1in}}
\providecommand{\U}[1]{\protect\rule{.1in}{.1in}}
\providecommand{\U}[1]{\protect\rule{.1in}{.1in}}
\providecommand{\U}[1]{\protect\rule{.1in}{.1in}}
\providecommand{\U}[1]{\protect\rule{.1in}{.1in}}
\providecommand{\U}[1]{\protect\rule{.1in}{.1in}}

\input{tcilatex}

\begin{document}

\title{Noncommutative magnetic moment, fundamental length and lepton size}
\author{{\large T. C. Adorno}\thanks{%
adorno@dfn.if.usp.br} {\large ,~~D. M. Gitman}\thanks{%
gitman@dfn.if.usp.br} \\
%EndAName
\textit{Instituto de F\'{\i}sica, Universidade de S\~{a}o Paulo,}\\
\textit{Caixa Postal 66318, CEP 05508-090, S\~{a}o Paulo, S.P., Brazil,}\\
{}\\
{\large A.~E.~Shabad}\thanks{%
shabad@lpi.ru}\\
\textit{P.~N.~Lebedev Physics Institute,}\\
\textit{Leninsky Prospekt 53, Moscow 117924, Russia}}
\maketitle
\date{}

\begin{abstract}
Upper bounds on fundamental length are discussed that follow from the fact
that a magnetic moment is inherent in a charged particle in noncommutative
(NC) electrodynamics. The strongest result thus obtained for the fundamental
lenth is still larger than the estimate of electron or muon size achieved
following the Brodsky-Drell and Dehlmet approach to lepton compositeness.
This means that NC electrodynamics cannot alone explain the whole existing
descrepancy between the theoretical and experimental values of the muon
magnetic moment. On the contrary, as measurements and calculations are
further improved, the fundamental length estimate based on electron data may
go down to match its compositeness radius.
\end{abstract}

\section{Introduction}

\bigskip

\subsection{The problem. Results and conclusions}

The clue issue in checking quantum electrodynamics (QED) is the measurement
of the magnetic moment of electron with the subsequent comparison of its
measured value with the anomalous electron magnetic moment calculated via
the Standard Model that is mostly QED in this case. Up to now, within every
experimental and theoretical accuracy achieved, these two values do
coincide. The allowed, within the errors, discrepancy between the
experimental and theoretical values of the electron magnetic moment is
expected to be diminishing on and on, as the precision grows, and hopefully
the coincidence between them will be maintained with better and better
accuracy. On the other hand, as far as one is seeking for possible
theoretical amendments to the Standard Model, admissible within the above
situation, one should confine their impact on the electron magnetic moment
to lie within the present experimental and theoretical indeterminacy. A
certain candidate for going beyond the standard QED is proposed by the
noncommutative (NC) electrodynamics. It was found recently \cite{AGSV} that
in the framework of that theory a static classical charge at rest \ carries
a magnetic moment, called NC magnetic moment, whose smallness is determined
by a noncommutativity parameter $\theta,$ supplying the theory with the
fundamental length \footnote{%
The noncommutativity by no means is the only method of introducing the
fundamental, or elementary, length into a theory. Throughout this paper,
however, we shall mean namely NC fundamental length when using this notion.
On the other hand, we do not know whether the fundamental length as it is
proposed by the noncommutativity mechanism is universal for all particles
and fields. For this reason we shall discuss its values independently when
we deal with different particles.} $l=\sqrt{\theta}$. By demanding that, for
the electron, the NC magnetic moment be less than the existing error in
measuring the electron magnetic moment we get an estimate from above on the
parameter $\theta$ and the associated fundamental length $l$ . Certain
restrictions on the fundamental length inherent in the NC theory also follow
from the existence of the NC magnetic moment of heavier charged particles.
However, the consideration of the noncommutative magnetic moment of the
proton and of its contribution to the hyperfine splitting of the energy
level $1S_{1/2}$ in a hydrogen atom did not lead \cite{AGSV} to any
essentially new estimate for the maximum fundamental length. On the
contrary, consideration of leptons did.

Once the NC magnetic moment is found to be inversely proportional to the
size of the electric charge, an important role in getting this estimate is
played by the size attributed to an electron, the smaller the size, the
larger the NC magnetic moment, the smaller the upper estimate on the NC
parameter and the fundamental length. We probe different assumptions
concerning the \textquotedblleft electron size\textquotedblright , the
ultimate one being that it is restricted from below only by the fundamental
length $l$ itself, since neither object is supposed to be smaller than it.
In this way a hitherto lowest upper bound on the fundamental length, as it
appears in the noncommutative field theory, was achieved in \cite{AGSV}. On
the other hand, after we update the famous electron size estimate\ due to
Brodsky-Drell-Dehlmet \cite{BrodDrell}, \cite{Dehmelt} (not based on any
noncommutative mechanism, but only on a consideration of a possible
compositeness, or divisibility of the electron) by taking into account the
most recent measurements of the electron magnetic moment, we find the
electron size results to be two orders of magnitude smaller than the boldest
estimate of the fundamental length obtained from speculations on
noncommutative magnetic moment. As far as in an NC field theory no size of
any physical object is admitted to be smaller than the fundamental length,
this means that no more than 1/100 part of the existing indeterminacy in the
knowledge of the electron magnetic moment may be at the best ascribed to the
contribution of the noncomutative magnetic moment. Then, two options remain.
Either there should be an extra extension beyond the standard QED, other
than NC electrodynamics, that may take responsibility for the main part of
the admitted, if any, deviation of the magnetic moment from the QED result,
or, what is more probable, this admitted deviation will be essentially
reduced by further more precise measurements.

The same analysis is repeated in the paper as applied to the $\mu $-meson.
The crucial difference with the electron case is that the difference between
the theoretical and experimental values of the muon magnetic moment exceeds
the limits admitted by the errors. So, no further technical advancement is
expected to be able to remove this contradiction, and our results make us
more definite in claming that the noncommutativity cannot provide for the
missing part of the muon magnetic moment, a different way for extending the
Standard Model remaining to be needed.

\bigskip

\subsection{Noncommutative magnetic moment}

In \cite{AGSV}, classical field equations in $U(1)_{\star}$-theory
(noncommutative electrodynamics) were formulated that -- at least within the
first order in the noncommutativity parameter $\theta$ -- restrain the gauge
invariance in spite of the presence of external current, known to violate it
(at least off-shell). By solving these equations electromagnetic field
produced by a finite-size static electric charge was found, and the fact
that this charge possesses a magnetic moment depending on its size was
established. Let the external current in the field equations of NC
electrodynamics be just a static electric charge distributed inside a sphere
of a finite radius $a$ with the uniform charge density 
\begin{equation*}
\rho\left( \mathbf{r}\right) =\frac{3}{4\pi}\frac{Ze}{a^{3}}\,,\ \ r<a,\quad
r=|\mathbf{r}|\,.
\end{equation*}
Outside the sphere there is no charge: $\rho\left( \mathbf{r}\right) =0,$ if 
$r>a$. The above finite-size static total charge $Ze,$ where $e$ is the
fundamental charge, produces not only the electrostatic field, but also
behaves itself as a magnetic dipole with the magnetic field given in the
remote region $r\gg a$ by the following vector-potential%
\begin{equation}
\hspace{0cm}\mathbf{A}=\frac{\left[ \mathbf{M}\times\mathbf{r}\right] }{r^{3}%
},\ \ \mathbf{M}=\boldsymbol{\theta}(Ze)^{2}\frac{2e}{5a}\,,
\label{magnmoment}
\end{equation}
where $\mathbf{M}$ was called NC magnetic moment of the charged particle.
Here the three spacial components of the vector $\boldsymbol{\theta}$ are
defined as $\theta^{i}\equiv(1/2)\varepsilon^{ijk}\theta^{jk},$ $i,j,k=1,2,3$
in terms of the antisymmetric noncommutativity tensor $\theta^{\mu\nu}$ that
fixes the commutation relations between the operator-valued coordinate
components $[X^{\mu},X^{\nu}]=i\theta^{\mu\nu},$ and only the space-space
noncommutativity, i.e. the special case of $\theta^{0\nu}=0$ in a certain
Lorentz frame, was considered.

The extension (size) $a$ of the charge in (\ref{magnmoment}) should be kept
nonzero in the spirit of NC theory that does not admit objects with their
size smaller than the fundamental length $l=\sqrt{\theta}$, where $\theta =|%
\boldsymbol{\theta}|$. For a point charge a magnetic solution also exists 
\cite{Stern}, although in this case it is not a magnetic dipole one. What is
more important is that that solution is too singular in the point $r=0$,
where the charge is located, and hence it cannot be given a mathematical
sense in terms of the distribution theory in a conventional way.

If we understand the radius $a$ in (\ref{magnmoment}) as the size of an
electrically charged fundamental particle ($Z=1$), we can speculate on what
the contribution of the noncommutativity into its magnetic moment $\mathbf{M}
$ may be. Certainly, this is expected to be very small, because of the
extreme smallness of the noncommutativity parameter $\theta $. It is
primarily supposed \cite{DFR} that the corresponding length $l=\sqrt{\theta }
$ should be of the Plank scale of $l\sim 10^{-33}$ cm (or $\Lambda _{\mathrm{%
Pl}}\sim 4\cdot 10^{19}$ Gev in energy units). The reason is that at so
small distances unification of gravity with quantum mechanics requires
quantization of space-time. Although the Plank scale is far beyond any
experimental reach, the everlasting problem is to estimate the upper limits
on $\theta $ basing on the existing and advancing experimental preciseness.
In \cite{AGSV} it was discussed what new restrictions on the extent of
noncommutativity may follow from the newly established fact that a charged
fundamental particle is a carrier of the magnetic moment (\ref{magnmoment})
in an NC theory, irrespective of its orbital momentum or spin. In the
present article we shall further elaborate this matter addressing the
charged leptons $e$ and $\mu $ as the \textquotedblleft
smallest\textquotedblright\ -- and hence providing the maximum contribution
of (\ref{magnmoment}) -- particles, to leave alone quarks -- also small, but
whose magnetic moment is beyond measurements.

\section{Upper bounds for fundamental length from noncommutative magnetic
moment}

\subsection{Limitations based on high-energy scattering estimates of lepton
sizes}

In high-energy electron-positron collisions leptons manifest themselves as
structureless particles (see e.g. \cite{BrodDrell} for an early discussion
of this point), described by a fundamental (local), not composite field. No
deviation from this rule has been up to now reported. Taking the LEP scale
of 200 Gev as an upper limit, to which this statement may be thought of as
checked, we must accept that a possible structureness of these leptons is
below the length (call it divisibility length) $r_{0}=10^{-3}Fm$. In our
further consideration we identify the charge extension $a$ with the
divisibility length, because it is hard to imagine a region occupied by a
charge that extends above this length, but cannot be divided into parts. (If
it could, either the resulting charge would acquire a continuous value,
smaller than $e$, which contradicts basic assumptions, or the resulting
charge would occupy a smaller space and we would be left again with smaller $%
a$, down to the divisibility length.)

\subsubsection{Electron}

Bearing in mind that, for electron, the existing local theory perfectly
explains the value of its magnetic moment $M_{\mathrm{e}}$, we expect that
the noncommutativity might only contribute into the experimental and
theoretical uncertainty $\delta M_{\mathrm{e}}$ existing in measuring and
calculating this quantity. A recent direct measurement of the anomalous
magnetic moment of electron, using the magnetic resonance spectroscopy of an
individual electron in the Penning trap \cite{Dehmelt}, gives the result 
\cite{Hanneke}, \cite{nakamura}%
\begin{equation}
\left. \left( \frac{M_{\mathrm{e}}}{\mu }-1\right) \right\vert _{\mathrm{MRS}%
}=0.00115965218073\pm 28\cdot 10^{-14},  \label{hanneke}
\end{equation}%
where $\mu =e/2m$ is the Bohr magneton. On the other hand, a new report \cite%
{Bouchendira} appeared on an \textit{independent} experimental determination
of the same magnetic moment with a matching accuracy, obtained with the use
of a measurement of the ratio $h/m_{\mathrm{Rb}}$ between the Plank constant
and the mass of the $^{87}$Rb atom. The result is 
\begin{equation}
\left( \left. \frac{M_{\mathrm{e}}}{\mu }-1\right) \right\vert _{\mathrm{Rb}%
}=0.00115965218113\pm 84\cdot 10^{-14}.  \label{theor}
\end{equation}%
Authors of \cite{Bouchendira} fit the value of the fine structure constant $%
\alpha $ in such a way as to make (\ref{theor}) coincide with the
theoretical prediction for the electron anomalous magnetic moment,
calculated (see \cite{Mohr} for a review) with the accuracy, including QED
calculations up to $(\alpha /\pi )^{4}$, also electroweak and hadronic
contributions (this fit leads to the so far most precise value $\alpha
^{-1}=137.035999037(91)$). For this reason the value (\ref{theor}) is
referred to as \textquotedblleft theoretical\textquotedblright . (Certainly,
the roles of (\ref{theor}) and (\ref{hanneke}) might be reversed.) The
theoretical, (\ref{theor}), and experimental, (\ref{hanneke}), values of the
electron magnetic moment do not contradict each other, demonstrating the
hitherto best confirmation of QED. The discrepancy between them%
\begin{equation}
\frac{\delta M_{\mathrm{e}}}{\mu }\sim 10^{-12}  \label{delta}
\end{equation}%
lies within the accuracy of measurements and calculations. We demand that a
possible contribution of the noncommutative magnetic moment in (\ref%
{magnmoment}) should not exceed it:%
\begin{equation}
\frac{\delta M_{\mathrm{e}}}{\mu }>\alpha \theta \frac{4m}{5a}\,,\ \alpha
=e^{2}\,.  \label{delM}
\end{equation}%
With the high-energy restriction on the size $a<r_{0}$ accepted above, Eq.(%
\ref{delM}) implies $\theta <\frac{\delta M_{\mathrm{e}}}{\mu }(5r_{\mathrm{0%
}}/4m\alpha )$. As $r_{0}\sim 10^{-3}Fm,$ we get from here and from\ (\ref%
{delta})\ the restriction on the fundamental length $l=\sqrt{\theta }<7\cdot
10^{-6}Fm=(28~Tev)^{-1}.$

\subsubsection{Muon}

The matters stand differently with the $\mu $-meson. In the literature, its
anomalous magnetic moment is calculated via the Standard Model with the
inclusion of the QED lowest-order $\mu $-$\gamma $ vertex, $Z$-boson,
neutrino and hadron lines. The deviation of the measured magnetic moment $\
M_{\mu \text{ }}$from the result of calculations makes the value (see A.
Hocker's and W.J. Marciano's 2009 update in \cite{nakamura}, also \cite{Mohr}
for a later detailed account) 
\begin{equation}
\frac{\delta M_{\mu }}{\mu }\simeq 25\cdot 10^{-10}.  \label{deltamu}
\end{equation}%
This exceeds about 3.2 times the estimated 1$\sigma $ error \cite{nakamura}.
It is believed that this discrepancy may be overcome by including
supersymmetry for amending the theoretical result. If, on the contrary, we
try to explain this discrepancy by the effect of NC magnetic moment of the
muon, we get in the way similar to the one described above in this
Subsection, using (\ref{deltamu}) and the same indevisability length $%
r_{0}\sim 10^{-3}Fm$ that $l$ is smaller than $2.8\cdot 10^{-5}Fm=$ (7 $Tev$)%
$^{-1}$ as the high-energy based estimate.

\subsection{Ultimate estimates}

Once there is no evidence for any electron extension, it is worth admitting
that it may be only restricted by the fundamental length. Then, using $a=l=%
\sqrt{\theta}$ in (\ref{delM}) and the indeterminacy (\ref{delta}), we
obtain the ultimate bound of $l<6.6\cdot10^{-8}Fm=(3\cdot10^{3}$ $Tev)^{-1}$%
. Dealing with the muon in the same way, but referring to (\ref{deltamu})
instead of (\ref{delta}), we obtain the ultimate estimate of $8\cdot
10^{-7}Fm=(240$ $Tev)^{-1}$.

\section{Upper bounds on fundamental length versus compositeness sizes of
leptons}

There are \cite{BrodDrell} much stronger restrictions on the lepton sizes
than those following from the high-energy collision experiments. These
extend to the energy scale far exceeding the accelerator means. The point is
that if one imagines a lepton as a bound state of much heavier particles so
that the binding energy compensates the most part of their masses to make
the resulting state light, the Bohr radius $R$ of the composite state -- to
be treated as its size -- would be much smaller than the Compton length of
the lepton $\lambda_{\mathrm{C}}$. According to the Drell-Hearn-Gerasimov
sum rule (see \cite{BrodDrell} for references) the deviation of the
anomalous magnetic moment $(M/\mu-1)$ from its QED value is proportional to
the ratio $R/\lambda_{\mathrm{C}},$which is the measure\ of compositeness.
Based on the experimental data on magnetic moments of the known composite
particles -- proton and triton -- plotted against their measured sizes, a
conjecture was formulated by Dehmelt \cite{Dehmelt} that the proportionality
coefficient should be of the order of unity. Then, $R=\lambda_{\mathrm{C}%
}\delta M/\mu$.

\subsection{Electron}

Referring to Eqs. (\ref{hanneke}, \ref{theor}) and using (\ref{delta}) we
may update Dehmelt's 1988 result for the electron of $R<4\cdot10^{-8}Fm$ to $%
R<4\cdot10^{-10}Fm$. This is two orders of magnitude smaller than our
ultimate estimate of $6.6\times10^{-8}Fm$ for the fundamental length
obtained in Subsection \textit{2.2}. (The use of the assertion $R=\lambda _{%
\mathrm{C}}\delta M/\mu$ together with (\ref{delM}) would result in the
condition $l<\sqrt{5/8\alpha}(\delta M/\mu)\lambda_{\mathrm{C}}=9.25R$,
weaker than the already accepted condition that the fundamental length \
should be smaller than any size, including the composite electron radius $R$%
, that is to $l<R$ $.$) So, Dehmelt's conjecture provides a stronger bound
on the fundamental length, than the noncommutative magnetic moment. This
means that no more than $10^{-2}$ part of the measured difference (\ref{delM}%
) may be at the most attributed to noncommutative contribution.

\subsection{Muon}

The muon radius estimated analogously, basing on the compositeness arguments
and on the theory-experiment discrepancy (\ref{deltamu}), gives the result $%
R_{\mu }\simeq 0.5\cdot 10^{-8}Fm.$ This is smaller than the ultimate
estimate of Subsection 2.2 based on muon data. Again, once the muon size
cannot be smaller than the fundamental length, this result indicates that
the NC magnetic moment alone definitely cannot take on the responsibility
for the discrepancy (\ref{deltamu}) between the theory and experiment, and
hence deviations from the Standard Model other than NC electrodynamics are
needed. Unlike the electron case above, one cannot set one's hopes upon the
future growth of precession of measurements to abandon this conclusion.

\section*{Acknowledgements}

T. Adorno thanks FAPESP, D. Gitman thanks CNPq and FAPESP for permanent
support, A. Shabad thanks FAPESP for support and USP for the kind
hospitality extended to him during the fulfillment of this work. A. Shabad
also acknowledges the support by RFBR under the Project 11-02-00685-a.

\end{document}